\documentstyle[a4wide,12pt,epsf]{article}
\begin{document}
\begin{flushright}
\baselineskip=12pt
{CERN-TH/2000-094}\\
{SUSX-TH-00-005}\\
{RHCPP99-08T}\\
{hep-th/0004052}\\
{March 2000}
\end{flushright}

\begin{center}
{\LARGE \bf CP violation including universal one-loop corrections and heterotic 
M-theory \\}
\vglue 0.35cm
{D.BAILIN$^{\clubsuit}$ \footnote
{D.Bailin@sussex.ac.uk \ Permanent address: Centre for Theoretical Physics, University of Sussex, Brighton BN1 9QJ, U.K.},
 G. V. KRANIOTIS$^{\clubsuit}$ \footnote
 {G.Kraniotis@rhbnc.ac.uk \ Permanent address: Centre for Particle Physics,  Royal Holloway and Bedford New College, University of London, Egham, Surrey TW20 0EX, U.K. } and A. LOVE$^{\spadesuit}$ \\}
        {$\clubsuit$ \it  Theory Division, CERN,\\}
{\it CH1211 Geneva 23, Switzerland. \\}
{$\spadesuit$ \it  Centre for Particle Physics, \\}
{\it Royal Holloway and Bedford New College, \\}
{\it  University of London, Egham, \\}
{\it Surrey TW20-0EX, U.K. \\}
\baselineskip=12pt

\vglue 0.25cm
ABSTRACT
\end{center}

{\rightskip=3pc
\leftskip=3pc
\noindent
\baselineskip=20pt
CP violation by soft supersymmetry-breaking terms in 
orbifold compactifications is investigated. We include the 
universal part of the moduli-dependent threshold corrections in the 
construction of the non-perturbative effective potential due to 
gaugino-condensation.
This allows interpolation of the magnitude of CP violating phases 
between the weakly and strongly coupled regimes. We find that the universal 
threshold corrections have a large effect on the CP violating phases in the 
weakly coupled regime.}

\vfill\eject
\setcounter{page}{1}
\pagestyle{plain}
\baselineskip=14pt

Too large CP violation by soft supersymmetry-breaking terms is 
a generic problem in supergravity and superstring theories.
This can result in a neutron electric dipole moment much larger 
than the experimental upper bound. We found elsewhere \cite{US} that the 
modular properties of orbifold compactifications of the weakly 
coupled heterotic string can lead to very small or zero CP 
violating phases in the soft supersymmetry-breaking terms even 
when the moduli have large phases at the minimum of the effective 
potential.

Recently, many new facts about the strong coupling limit 
of string theory have been accumulated with the advent of M-theory
\cite{HORWIT}-\cite{MUNOZ}. 
One would like to interpolate between the weak and strong coupling 
limits of this 
unique theory and study differences (if any) in the resulting 
phenomenology. As was emphasized first by Nilles and Stieberger 
\cite{NILLES}, the calculated gauge-group-independent 
universal threshold effects 
in the gauge kinetic function of weakly coupled orbifold theories allow 
such an interpolation to be made for some purposes. 

As we shall see in this paper, the magnitude of CP violation 
due to soft supersymmetry-breaking terms 
varies as 
one goes between regions corresponding to weakly and strongly coupled 
regimes of the moduli space.
We shall also see that the universal threshold corrections have a large 
effect on the CP violating phases in the weakly coupled regime.

The gauge kinetic function including universal 
threshold effects  is given by
\begin{equation}
f_a=S-\frac{1}{16\pi^2}\sum_i\frac{|G_i|}{|G|}\left[(b_a^{N=2})_i
\ln (\eta(T_i))^4-\sigma_1(T_i,U_i)\right]
\label{kinefu}
\end{equation}
where $|G_i|$ are the orders of the  subgroups of the point group $G$ 
which leave the $i-$th complex plane unrotated in the six compact 
dimensions. Also, for ${\rm Re}T>{\rm Re}U$
 \cite{NILLES} 
\begin{eqnarray}
\sigma_1(T,U)
&=&-2 {\rm ln}[j(T)-j(U)] \nonumber 
\\
&-&2 \sum_{(k,l)>0}\;c(kl)\; kl\; {\rm ln}[1-e^{-2\pi (kT+lU)}]
\end{eqnarray}
The notation $(k,l)>0$, means 
that we sum over the orbits: (i) $k>0, \l=0$, (ii) $k=0,\ l>0$, (iii) $k,l>0$, (iv) $k=1,\l=-1$
The  coefficients $c(n)$ are defined by
$F(q)=\sum_{n=-1}^{\infty} c(n)q^n=\frac{E_4 E_6}{\eta^{24}}$, where 
$E_4, E_6$ are the Eisenstein series with modular weight 4 and 6 respectively. 
The term  $j(T)-j(U)$ is the 
denominator formula of the Monster Lie Algebra \cite{RICHARD}.

The 
non-perturbative 
superpotential due to a single gaugino condensate is taken to be
$W_{np} \sim e^{\frac{24\pi^2}{b_a}f_a}$,
and substituting (\ref{kinefu}) 
gives 
\begin{equation}
W_{np}\sim e^{\frac{24 \pi^2}{b_a} S}\prod_{i} e^{\frac
{3\sigma_1(T_i,U_i)}{2 b_a}\frac{|G_i|}{|G|}}\prod_i \Bigl(
\eta(T_i)\bigr)^{\frac{-6(b_a^{N=2})_i}{b_a}\frac{|G_i|}{|G|}}
\end{equation}
The K\"{a}hler potential, including non-perturbative 
corrections to the dilaton part of the K\"{a}hler potential parametrized 
by the function $P(y)$, is given by 
\begin{equation}
K=P(y)+\hat{K}
\end{equation}
where $\hat{K}=-\sum_{i}{\rm ln}(T_i+\bar{T}_i)$. 
The modular invariant function $y$ includes loop-corrections due to universal 
threshold corrections and is given by the equation
\begin{equation}
y=S+\bar{S}-\Delta
\end{equation}
where \cite{NILLES}
\begin{equation}
\Delta=\sum_{i}\frac{\delta^i_{GS}}{8 \pi^2}ln (T_i+\bar{T}_i)+
\frac{1}{8\pi^2}\sum_{i}\frac{|G_i|}{|G|}G^{(1)}(T_i,U_i,\bar{T}_i, \bar{U}_i)
\label{KAKA}
\end{equation}
with $\delta^i_{GS}$ the Green-Schwarz anomaly cancelling coefficients, and \cite{NILLES,
RIZOS}
\begin{eqnarray}
G^{(1)}(T,U,\bar{T},\bar{U})&=&-\frac{4\pi}{3}\frac{({\rm Re}U)^2}{{\rm Re} T}\Theta({\rm Re} T-{\rm Re} U) \nonumber \\
&-&\frac{4\pi}{3}\frac{({\rm Re} T)^2}{{\rm Re} U}\Theta({\rm Re} U-{\rm Re} T) \nonumber \\
&+&\frac{1}{\pi {\rm Re} T {\rm Re} U} {\rm{Re}}\tilde{P}(T-U) \nonumber \\
&-&\frac{60}{\pi^2 Re T Re U}(\zeta(3)+4\pi {\rm{Re}}\sum_{k>0}
\tilde{P}(kT)+4\pi {\rm Re}\sum_{l>0}\tilde{P}(l U)) \nonumber \\
&+&{\rm Re} \sum_{k,l>0} \frac{c(kl)}{\pi {\rm Re} T {\rm Re} U}\tilde{P}(kT+lU)
\end{eqnarray}
The function  ${\tilde{P}}(x)$ is defined by ${ 
\tilde{P}}(x)={\rm Re} x {\cal L}i_2(e^{-2\pi x})+
\frac{1}{2\pi} {\cal L}i_3(e^{-2\pi x})$
where ${\cal L}i_j$ are the polylogarithms. 
The polylogarithms are given in the unit disk $|z|<1$ by the expression
\begin{eqnarray}
{\cal L}i_1(z)&=&\sum_{n=1}^{\infty}\frac{z^n}{n} \nonumber \\
{\cal L}i_2(z)&=&\sum_{n=1}^{\infty}\frac{z^n}{n^2}  \nonumber \\
{\cal L}i_3(z)&=&\sum_{n=1}^{\infty}\frac{z^n}{n^3} \nonumber \\
\end{eqnarray}
The dilogarithm ${\cal L}i_2$ and 
the trilogarithm ${\cal L}i_3$ can be continued 
analytically outside the unit circle, i.e. \cite{Lewin}
\begin{eqnarray}
{\cal L}i_3(e^{x})&=&{\cal L}i_3(e^{-x})+
\frac{\pi^2}{3} x-\frac{i \pi}{2}x^2 -\frac{1}{6}x^3 \nonumber \\
{\cal L}i_2(e^{x})&=&-{\cal L}i_2(e^{-x})+\frac{\pi^2}{3}-i\pi x-\frac{1}{2}x^2
\label{anal}
\end{eqnarray}
The $L$-functions obtained by constructing series of polylogarithms with 
Fourier coefficients the Fourier expansion coefficients of the modular 
form  $\frac{E_4 E_6}{\eta^{24}}$ have highly non-trivial modular properties.
For instance under $U\rightarrow 1/U,T\rightarrow T$

\begin{eqnarray}
\sum_{r>0}c(kl){\cal L}i_3(e^{-2\pi (kT+lU^{'})})&=&-U^{-2}\sum_{r>0}c(kl)
{\cal L}i_3(e^{-2\pi (kT+lU)}) \nonumber \\
&+&-\frac{1}{2}\zeta(3)c(0)
(U^{-2}+1)+ \nonumber \\
&-&\frac{4\pi^3}{3}(U+5U^{-1}+U^{-3})
\end{eqnarray}
where $U^{'}=1/U$ and we are in the chamber ${\rm Re}T>{\rm Re}U,\ {\rm Re}U^{'}$.
We have verified this numerically as a check on the accuracy of our codes. We have also verified numerically that the prepotential satisfies the identity (4.36) and (4.37) of Harvey and Moore \cite{HM}.

The effective potential in the case of a single 
modulus $T_3$, and $U$-modulus fixed at a constant value, 
is given by the expression
\begin{eqnarray}
e^{-G}\; V_{eff}
&=&-3+{\left(\frac{d^2 P}{dy^2}\right)}^{-1}{\Bigl| \frac{dP}{dy}+\frac{\partial{ln W_{np}}}
{\partial S}\Bigr|}^2 \nonumber \\
&+&{\left[ 1+\delta_3 \frac{dP}{dy}-\frac{1}{24 \pi^2}\frac{dP}{dy}
(T_3+\bar{T}_3)^2 G_{3\bar{3}}^{(1)}\right]}^{-1} \\
&\times&\Bigl|(T_3+\bar{T}_3)
\frac{\partial {\rm ln}W_{np}}{\partial T_3}-1+
\frac{\partial {\rm ln}W_{np}}{\partial S}\left(\delta_3+
(T_3+\bar{T}_3)\frac{1}{24\pi^2}G_3^{(1)}\right) \Bigr|^2
\end{eqnarray}
where $e^G=|W_{np}|^2 (T_3+\bar{T_3})^{-1}e^{P(y)}$, 
$\delta_3=\frac{\delta_3^{GS}}{8\pi^2}$,  and for a pure gauge hidden sector $\delta_3^{GS}=\frac{b_a}{3}\left( 1-2\frac{|G_3|}{|G|}\right)$; the suffices $3$ and $\bar{3}$ on $G^{(1)}$ denote differentiation with respect to $T_3$ and $\bar{T}_3$. To ensure physical dilaton kinetic energy terms we require $P''\equiv \frac{d^2 P}{dy^2} >0$. We have checked that $V_{eff}$ is invariant under the modular transformations $T \rightarrow T'=\frac{1}{T}$ and $T \rightarrow T+i$ for test values of $T$ and $U$ satisfying ${\rm Re}T, \ {\rm Re}T'<{\rm Re}U$. This requires the analytic continuation (\ref{anal}) of the polylogarithms to the region of moduli space reached by the above modular transformation. However, we were unable to obtain the modular repeats of the minima of $V_{eff}$ obtained later, with $U=e^{i\pi /6}$, because of the slow rate of convergence 
of the $L$-functions for these values. 

The soft supersymmetry-breaking $A$ terms are in general given by 

\begin{eqnarray}
A_{\alpha\beta\gamma}&=&{\left(\frac{d^2 P}{dy^2}\right)}^{-1} \left(
\frac{dP}{dy}+\frac{24\pi^2}{b}\right)\frac{dP}{dy} \nonumber \\
&+&C_{\bar{i}j}^{-1}\left(\Delta_{\bar{i}}\frac{\partial {\rm ln}\bar{W}_{np}}
{\partial \bar{S}} 
+\frac{\partial {\rm ln}\bar{W}_{np}}{\partial \bar{T}_i}-
(T_i+\bar{T}_i)^{-1}\right) \nonumber \\
&\times& \left(\frac{\partial log h_{\alpha\beta\gamma}}{
\partial T_j}-(T_j+\bar{T}_j)^{-1}[1+n_{\alpha}^{j}+
n_{\beta}^j+n_{\gamma}^j]\right)
\end{eqnarray}
where the superpotential term for the Yukawa couplings of 
$\phi_{\alpha},\phi_{\beta}$ and $\phi_{\gamma}$ is 
$h_{\alpha\beta\gamma}\phi_{\alpha}\phi_{\beta}\phi_{\gamma}$, the 
modular weights of these states are $n_{\alpha}^3,n_{\beta}^3$ and 
$n_{\gamma}^3$, and the usual rescaling by a factor 
$\frac{W_{np}}{|W_{np}|}$ required to get from the supergravity 
theory derived from the orbifold compactification of the 
superstring theory to the spontaneously broken  globally 
supersymmetric theory has been carried out.
In the case of a single modulus $T_3$
\begin{eqnarray}
C_{3\bar{3}}=(T_3+\bar{T}_3)^{-2}\left[1+
\delta_3\frac{dP}{dy}-\frac{1}{24 \pi^2}\frac{dP}{dy}
(T_3+\bar{T}_3)^2 G_{3\bar{3}}^{(1)}\right]
\end{eqnarray}
and 
\begin{eqnarray}
\Delta_3=
\delta_3 (T_3+\bar{T}_3)^{-1}+
\frac{1}{8\pi^2}\frac{|G_3|}{|G|}G_{3}^{(1)}
\end{eqnarray}
In particular, for the $Z_6^{'}$ orbifold with standard embedding, 
the hidden sector is a pure $E_8^{'}$ sector, so $b_a=-90$; also $\frac{|G_i|}{|G|}=\frac{1}{3}$, so  $\delta_3^{GS}=-10$; the 
U modulus is fixed at $e^{i\pi/6}$ and $W_{np}$ is given by 
\begin{equation}
W_{np} \sim e^{-\frac{4\pi^2}{15}S} 
e^{-\frac{\sigma_1(T_3,U_3)}{60}\frac{1}{3}}
{\eta(T_3)}^{-4/3}
\end{equation}
The case without inclusion of the universal terms is obtained by setting $\sigma_1= G^{(1)}=0$
The Yukawa couplings with a non-trivial 
moduli dependence are given by
\begin{eqnarray}
h(T_3,k)&\sim &e^{-\frac{2}{3}\pi k^2 T_3} \nonumber \\
&\times & [\theta_3 (i k T_3,2 i T_3) 
\theta_3 (i k T_3,6 i T_3) \nonumber \\
&+& \theta_2 (i k T_3, 2i T_3) \theta_2 (i k T_3,6 i T_3)]
\end{eqnarray}
where $k=0,\pm 1$ is related to the fixed points associated with the
 three (twisted-sector) states $\phi_{\alpha},\phi_{\beta}$ and $\phi_{\gamma}$.
These couplings are invariant under the modular transformation $T\rightarrow T+i$,
and  the modular weights are $n_{\alpha}^3=n_{\beta}^3=n_{\gamma}^3=-2/3$.
We do not consider the more model-dependent $B$ soft supersymmetry-breaking 
term.

\begin{figure}
\epsfxsize=6in
\epsfysize=8.5in
\epsffile{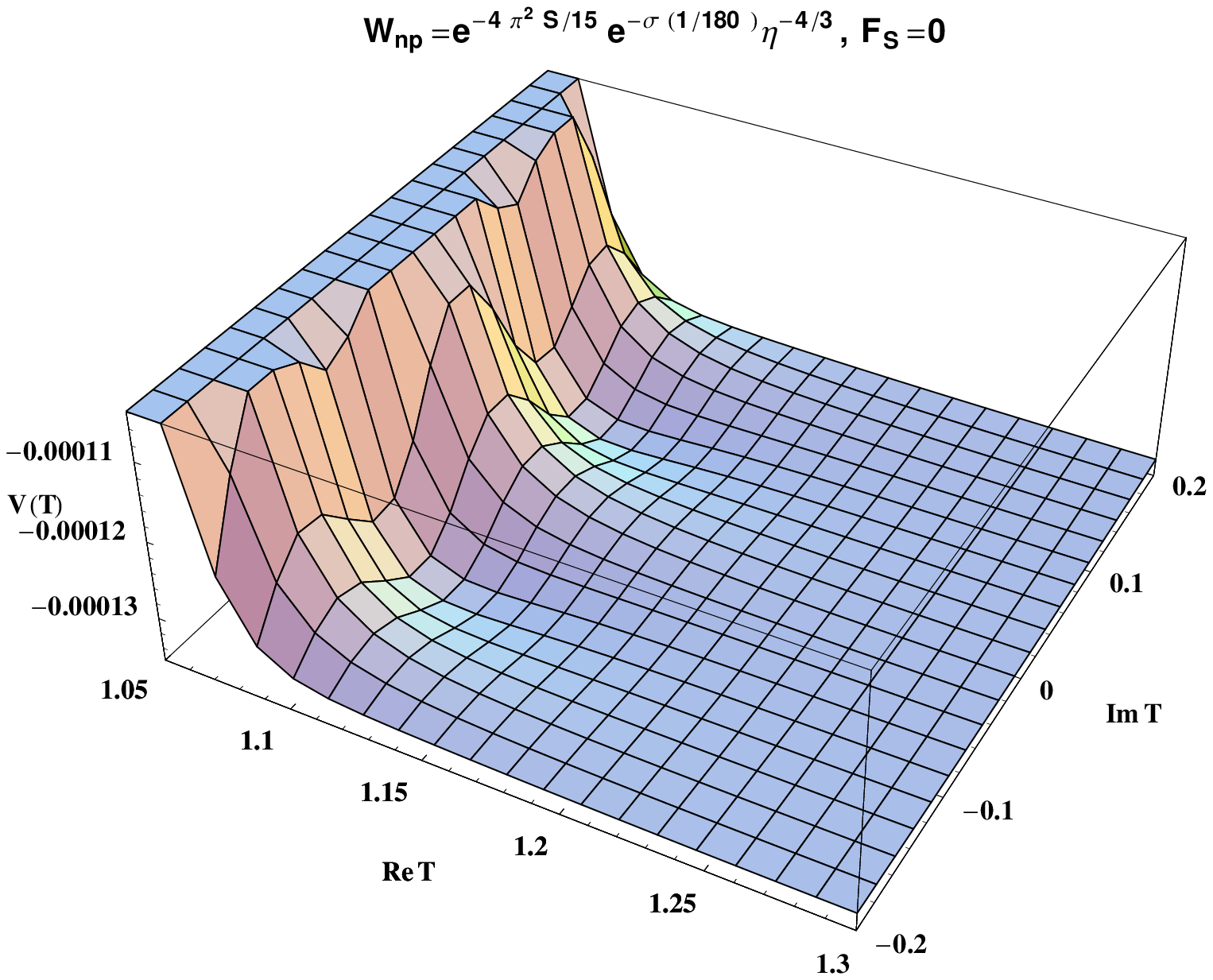}
\caption{Moduli potential in the limit $F_S=0$}
\label{modupot}
\end{figure}

\begin{figure}
\epsfxsize=6in
\epsfysize=8.5in
\epsffile{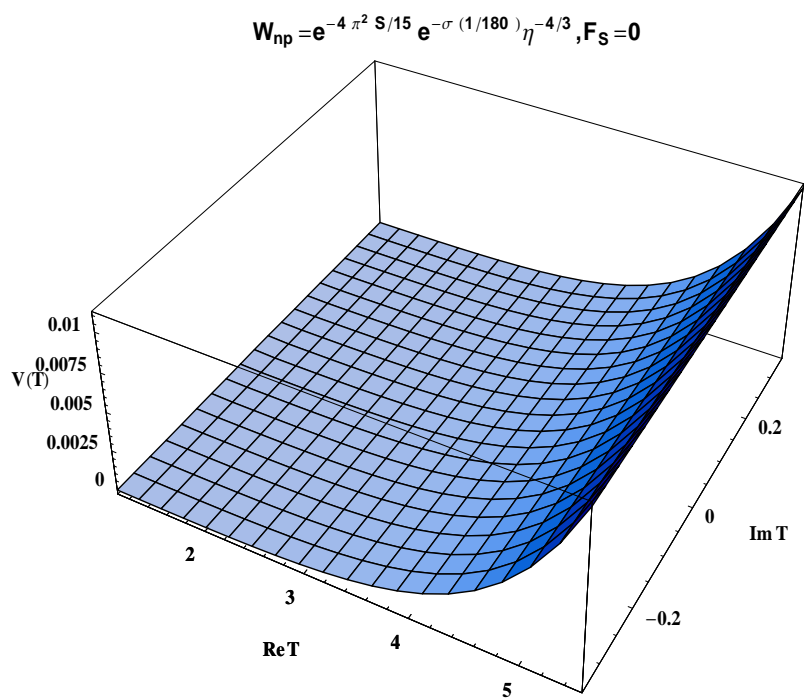}
\caption{Moduli potential in the limit $F_S=0$ for large $Re T$}
\label{modu2}
\end{figure}

Let us start 
our discussion with the moduli-dominated limit, i.e. we neglect the dilaton F-term
 $F_S \equiv P'+ W^{-1}W_S$; (then $P^{'}=-\frac{24\pi^2}{b}$, and  $P^{''}$ is
arbitrary.) In this case, minimising  $V_{eff}$ (see Figs.\ref{modupot}  
and \ref{modu2}) with respect to $T_3$ leads to 
\begin{equation}
T_3=1.1239\pm 0.0830\; i
\end{equation}
which is in the interior of the 
standard fundamental domain of $PSL(2,Z)$. $V_{eff}$ is negative at this point, and
the phase of the soft supersymmetry-breaking $A$ term
 is $\phi(A)=3.8\times 10^{-2}$.
This should be contrasted with previous results \cite{US}, obtained without inclusion of the universal threshold corrections, the only modular function present being the Dedekind $\eta$ function, in which {\em real} values of $T$ at the minimum were obtained. In the present case, omission of the universal terms gives a minimum at $T=1.33$.  However,
 the picture  is  
consistent with previous results \cite{US}
in which, when the absolute modular invariant $j(T)$ was  present 
in $W_{np}$, larger phases 
 occurred when the minima of $V_{eff}$ were at values of $T$ in the $interior$ of the 
standard fundamental domain of $PSL(2,Z)$ than for 
values of $T$ $on$ the boundary.

\begin{figure}
\epsfxsize=6.8in
\epsfysize=9.2in
\epsffile{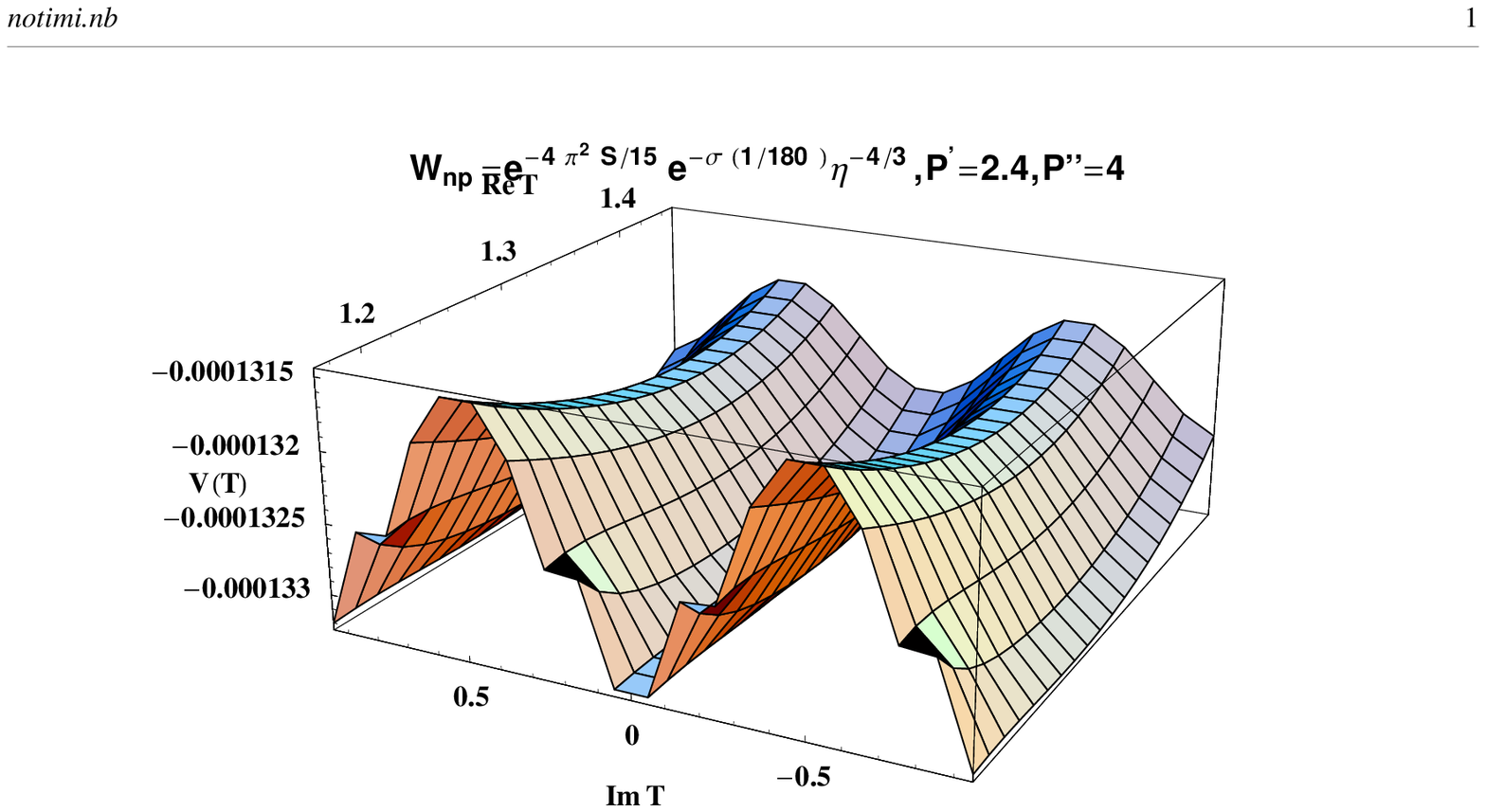}
\caption{Moduli potential for $P^{'}=2.4$ and $P^{''}=4$.}
\label{mod3}
\end{figure}

For $P'=1/3$ and $P''=1/4$ we also find the minimum in the interior of the fundamental domain, this time at
\begin{equation}
T_3=1.070 + 0.44986\; i
\end{equation}
which gives $\phi(A)=0.05$. Again, this differs considerably from the real minimum which occurs when the universal terms are absent.
If instead we consider $P^{'}(y)=3.4, P^{''}(y)=2$, the 
minimum of $V_{eff}$
is again in the interior of the fundamental domain, at
\begin{equation}
T_{3}=1.10046+0.23770\;i 
\end{equation}
and the phase of the trilinear soft-terms in this case is 
$\phi(A)={\rm O}(10^{-6})$. 
Other values of the non-perturbative K\"{a}hler potential parameters 
produce  minima at real $T$, but with significantly different values from the case when the universal terms are absent. For example, for $P^{'}=2.4, P^{''}=4$ (see Fig. 
\ref{mod3}), we find the minimum at
\begin{equation}
T_3=1.29459+n\;i
\end{equation}
Without universal threshold effects the minimum is located at 
$T_3=1.383$.


In conclusion, the introduction of universal threshold corrections in 
the construction of the gaugino condensate superpotential 
$W_{np}$ and the K$\rm{\ddot a}$hler potential $K$ 
leads to CP-violating phases in 
soft supersymmetry-breaking terms  of order $10^{-4}-10^{-2}$ in 
some regions of the 
parameter space close to the current experimental limit from the neutron 
dipole 
moment.
This is the case for the particular orbifold model studied in this paper, 
namely the $Z_6^{'}$ orbifold with single gaugino condensate and an $E_8^{'}$
pure gauge hidden sector.
In other regions of the parameter space CP violation is zero or 
negligible. 
Both the absolute modular invariant $j(T)$ and modular 
forms formed from the polylogarithms ${\cal L}i_m$ were present in the 
effective Lagrangian besides the Dedekind eta function.
This is contrast to the case without universal corrections where the 
standard form of $W_{np}$ contains only the Dedekind eta function. Then 
the CP violating phases were always zero or extremely small 
(of $O(10^{-15}$).

\begin{figure}
\epsfxsize=6.5in
\epsfysize=8.5in
\epsffile{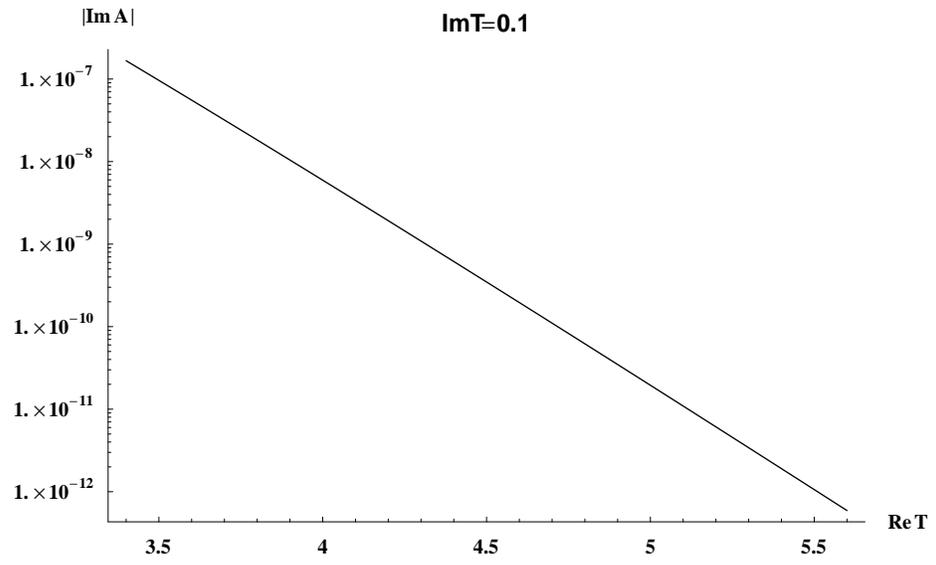}
\caption{Imaginary part of $A$ soft term as a function of ReT for ImT=0.1.}
\label{asoft}
\end{figure}

\begin{figure}
\epsfxsize=6in
\epsfysize=8.3in
\epsffile{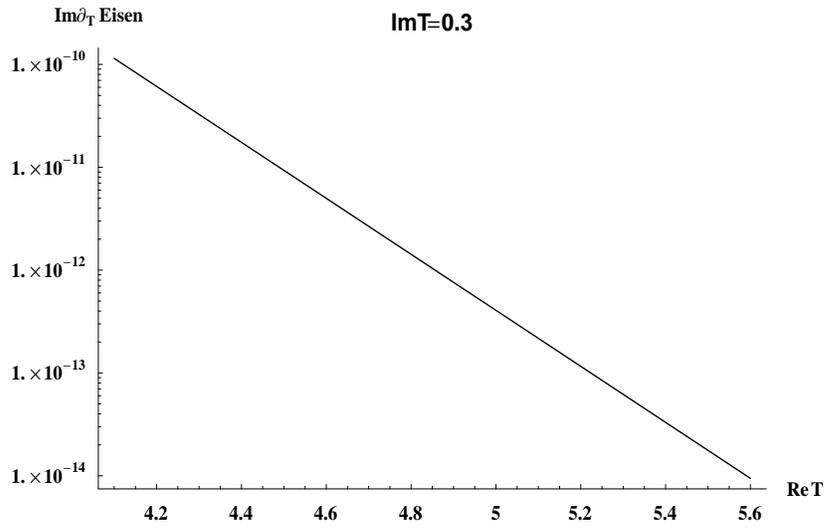}
\caption{Imaginary part of $Im\; Eisen$ as a function of ReT for ImT=0.3.}
\label{asoft3}
\end{figure}

Although minima with large values of $T$ were not obtained 
in our calculations, if such complex minima had been obtained 
(corresponding to strongly coupled M-theory), 
it is clear that the phases in the soft supersymmetry-breaking $A$
 terms would have been negligible owing  to the 
properties of the modular functions.  We studied the variation of $|{\rm Im}A|$ as a function of ${\rm Re}T$ for various values of 
${\rm Im}T$. One such case is displayed in Fig. \ref{asoft}, which shows 
that $|{\rm Im}A|\rightarrow 0$ rapidly for large  ${\rm Re}T$. This behaviour originates from the generalised  Eisenstein functions 
and Jacobi theta functions $\theta_i$ involved in the soft supersymmetry-breaking $A$ term. In Fig. \ref{asoft3} we plot the imaginary part of 
the derivative of the generalised Eisenstein function, $Eisen\equiv -\frac{2}{3}\hat{G}(T,\bar{T})-\frac{1}{90}\partial_T(G^{(1)}+\frac{1}{2}\sigma_1)$, as a function of ${\rm Re}T$; (As usual, $\hat{G}\equiv (T+\bar{T})^{-1}+2 \partial_T {\rm ln}\eta $.) Like $\hat{G}$, we again see that ${\rm Im}Eisen\rightarrow 0$ rapidly as ${\rm Re}T\rightarrow \infty$. The Jacobi theta functions  have a similar behaviour as  ${\rm Re}T\rightarrow \infty$.    


Thus the following physical picture seems 
to emerge. At weak coupling, we find minima in the interior of the fundamental 
domain, and the theory breaks CP. As we go to 
strong-coupling, i.e. large $T$-minimum, CP violation becomes negligible 
owing to properties of the modular functions.

There remains the following question. Is it possible to obtain  strong-coupling 
$M$-theory minima in modular invariant theories? 
With a single gaugino condensate, the answer is negative.
In Fig. 5 we plot the potential $V_{eff}$ (for $F_S=0$) as a function of Re$T$ and Im$T$, and
we observe that the potential does not have a 
minimum for large values of Re$T$.
It might be possible to obtain an M-theory minimum in modular theories 
using more than one gaugino condensate and/or using five-branes in the 
effective action. However, in the case of five-branes it is unclear how 
modular invariance can be incorporated.
On the other hand, even if one manages to obtain a minimum at large Re$T$, it  seems 
that CP violation in the soft supersymmetry-breaking terms will be 
negligible, 
in the strong coupling regime, due to properties of the 
Eisenstein and Jacobi modular functions.

\section*{Acknowledgements}
This research is supported in part by PPARC. DB and GVK thank the Theory Division at CERN for hospitality received during the completion of this work.

\newpage

\newpage

\newpage

\newpage

\end{document}